\documentclass[12pt]{article}
\usepackage{latexsym}
\usepackage{graphics}
\usepackage{epsfig}
\usepackage{epstopdf}
\usepackage{amsmath}
\usepackage{cite}
\usepackage{hyperref}

\begin{document}
\begin{center}
\textbf{\Large Energy redistribution in hierarchical systems of
oscillators} \vspace{0.5 cm}

Danylenko V.A., Mykulyak S.V.\footnote{e-mail:
\url{mykulyak@ukr.net}}, Skurativskyi S.I.\footnote{e-mail: \url{
skurserg@gmail.com}} \vspace{0.5 cm}

Subbotin institute of geophysics, Nat. Acad. of Sci. of Ukraine

     Bohdan Khmelnytskyi str. 63-G, Kyiv, Ukraine

\end{center}

\begin{quote} \textbf{Abstract.}{\small  The article deals with the mathematical
model for media with hierarchical structure. Using the hamiltonian
formalism, the dynamical system describing the state of
hierarchically connected structural elements  was derived.
According to the analysis of the Poincar\'{e} sections, we found
the localized quasi-periodic and chaotic trajectories in the
three-level hierarchical model. Moreover, studies of correlation
functions shown that the power spectrum for three-level model
possesses local maxima characterizing  temporal scales with strong
correlation. Using the Fourier analysis of the solution's
components, we have studied the distribution of energy injected in
the system over hierarchical levels. Dynamical phenomena in the
multi-level system were studied as well.}
\end{quote}

\vspace{0.5 cm}

\section{Introduction}\label{intro}
One productive way to study dynamic processes in the lithosphere
is an approach in which the lithosphere is considered as a
hierarchical system of volumes (blocks)
 \cite{Alexeevskaya,Sadovskii,Gabrielov86,Keilis-Borok,Gabrielov00},
where the block sizes are scaled in the interval from kilometers
(tectonic plates) to millimeters (grains of rocks). The main
features of dynamic behavior of such systems are high degree of
nonlinearity, complicated interaction between structural elements,
the redistribution of energy between hierarchical levels.  These
processes are manifested in the generation of high-frequency
oscillations during the application of low-frequency disturbances
in near - wellbore area \cite{Kurlenya00}, in the change of
spectra of aftershocks sequences  \cite{Sadovskii91}, in the
existence of a wide range of nonlinear waves after explosions in
rock massifs \cite{Kurlenya99} and so on. The exchange of energy
between hierarchical levels is the significant factor in the loss
of stability of block systems during seismic energy release
\cite{Starostenko96,Sadovskii87}. In order to study these
dynamical phenomena, we propose a model for media consisting of
the hierarchically connected elements (blocks). Most studies of
the dynamics of hierarchical systems of oscillators have been
focused on the synchronization processes
\cite{Winfree,Kuramoto75,Kuramoto84,Bonilla,Kori,Arenas,Ott08,Zhuo,Prignano,Skardal,Guo,Villegas},
instead we are interested in exploring exchange of energy between
hierarchical levels, in particular,  energy  transfer from the
upper level to the lower one.

The rest of this report is organized as follows. In
Sec.\ref{sec:1} we introduce the models describing the
hierarchically connected oscillators. Using the Poincar\'{e}
section technique, correlation and spectral analysis, in
Sec.\ref{sec:2} we  investigate the simplified model for
three-level system of oscillators.  Sec. \ref{sec:3} is devoted to
dynamical phenomena being observed in the multi-level hierarchical
system when model parameters are varied. Concluding remarks are
incorporated in final Sec.\ref{sec:4}.

\section{Models for hierarchical media}\label{sec:1}

\begin{figure}[h]
\begin{center}
\includegraphics[width=7 cm, height=6 cm]{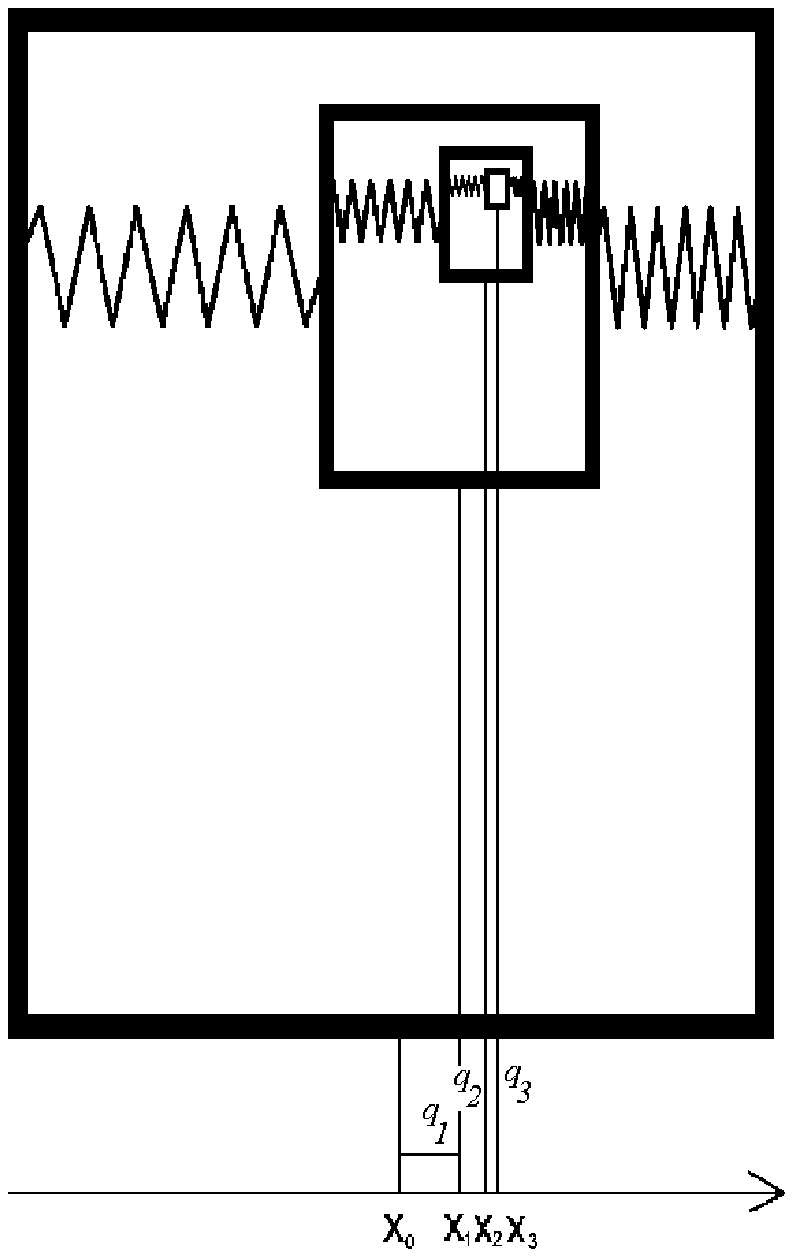}
\hspace{2 cm}
\includegraphics[width=6 cm, height=6 cm]{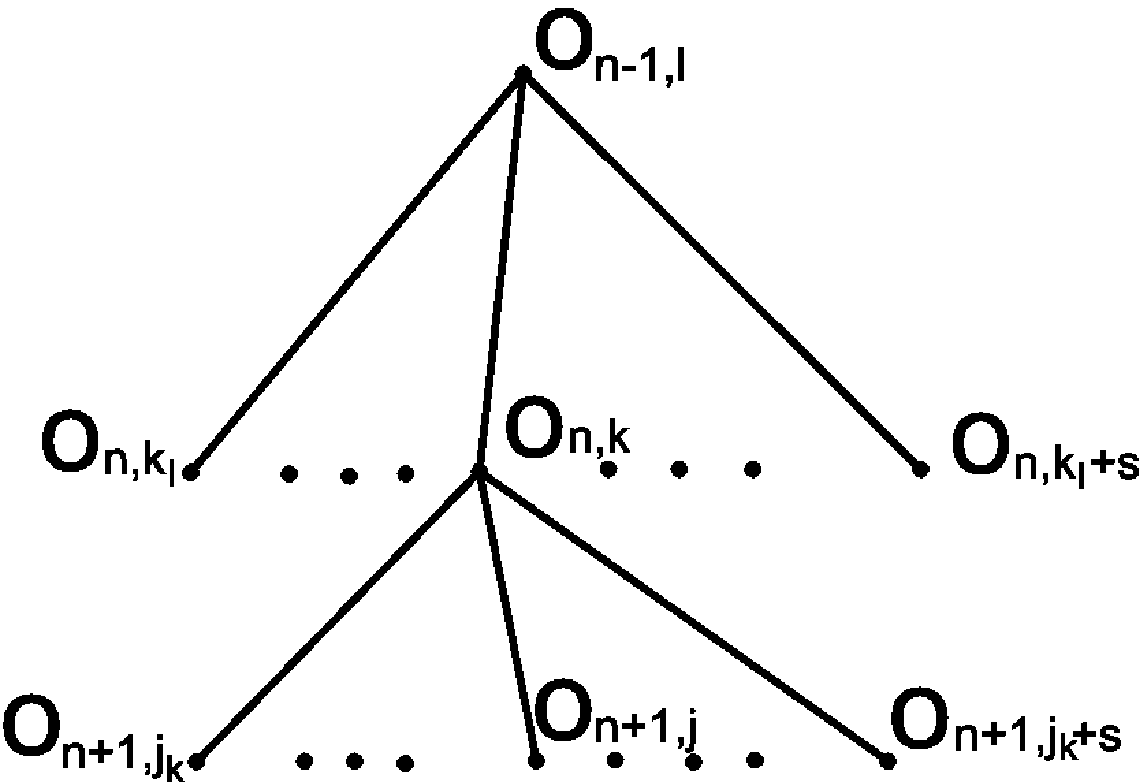}\vspace{0.5 cm}\\
a \hspace{8 cm} b
\end{center}
 \caption{The model of hierarchical medium (a)
and corresponding tree (b).}\label{Scheme}
\end{figure}

We model the hierarchical block medium by the embedded system of
anharmonic oscillators (Fig. \ref{Scheme}a). The part of
hierarchical system as a tree is shown in Fig.\ref{Scheme}b.

Assume that the model incorporates $N$ levels and each oscillator
from $n$th $(1<n<N)$ level connects with $s$ oscillators from the
$(n+1)$th level, except the lowest level oscillators. The
Hamiltonian of this system has the following form
\begin{equation}\label{lagrang}
H = \sum\limits_{n = 1}^N {\sum\limits_{k = 1}^{k_n } {\left(
{\frac{p_{nk}^2 }{2m_{nk}}  + \frac{C_{nk}}{\alpha} \left| {x_{nk}
- x_{(n - 1)\ell_k } } \right|^\alpha } \right)} },
\end{equation} where index
$(nk)$  denotes  the oscillator's position, namely, the $n$th
level and the $k$th place; $\ell_k$ is the place of oscillator
from the $(n-1)$th level connected with the $(nk)$th oscillator; $
m_{nk} $ and $p_{nk}$ are the mass and, respectively, the momentum
of the $(nk)$th oscillator; $x_{nk}$ is the oscillator's
coordinate; $C_{nk}$ is the stiffness of link, $\alpha$ is a
constant.

Inserting $H$ into the equations
     \[
\dot x_{nk}  = \frac{{\partial H}}{{\partial p_{nk} }}, \quad \dot
p_{nk} = - \frac{{\partial H}}{{\partial x_{nk} }},
\]
 one can obtain the following system of differential equations
\begin{equation}\label{model_ODE1}
\begin{split}
\dot x_{nk}  &= p_{nk} /m_{nk},\\ \dot p_{nk}  &=  - C_{nk} \left|
{x_{nk}  - x_{(n - 1)\ell_k} } \right|^{\alpha - 1}
\mbox{sgn}(x_{nk} - x_{(n - 1)\ell_k} )
\\&+\sum_j^{j+s}C_{(n+1)j}\left| {x_{(n+1)j} - x_{n k} }
\right|^{\alpha - 1}\mbox{sgn} (x_{(n+1)j} - x_{(n+1)k} ),
\end{split}
\end{equation}
where $\mbox{sgn}(x)=1$ if $x \geq 0$,  otherwise
$\mbox{sgn}(x)=-1$. Considering the oscillators, which are
identical on each level and move synchronously, system
(\ref{model_ODE1}) can be written as follows
\begin{equation}\label{model_ODE2}
\begin{split}
\ddot x_n  =  - \omega_n^2 \left| {x_n  - x_{n - 1} }
\right|^{\alpha  - 1}  \mbox{sgn} (x_n  - x_{n - 1} )
+\omega_{n+1}^2 s \frac{m_{n+1}}{m_n}\left| {x_{n+1}  - x_n}
\right|^{\alpha  - 1} \mbox{sgn} (x_{n+1}  - x_n ),
\end{split}
\end{equation}
where $\omega_n^2  = \frac{C_n }{m_n }$.

For convenience, let us introduce the new variables $q_n = x_n -
x_{n-1} $ having the meaning of  displacements from the steady
state, $x_0=\mbox{const}$. Then  $x_n=\sum_{s=1}^N q_s$. Finally,
system (\ref{model_ODE2}) reads
\begin{equation}\label{syst}
\begin{split}
\sum_{s=1}^N \ddot q_s  =  &- \omega_i^2 \left| q_i
\right|^{\beta} \mbox{sgn} (q_i )+\omega_{i+1}^2 s
\frac{m_{i+1}}{m_i}\left| {q_{i+1}} \right|^{\beta} \mbox{sgn}
(q_{i+1}  ).
\end{split}
\end{equation}
Using the designations  $F_i=\omega_i^2 \left| q_i
\right|^{\beta}\mbox{sgn}(q_i)$, $\varphi_i = s m_{i+1}/m_i$, we
present system (\ref{syst}) in the form
\begin{equation}\label{syst1}
\begin{split}
\ddot q_1=-F_1+\phi_1F_2,\\
\ddot q_n=F_{n-1}-F_n(1+\varphi_{n-1})+\varphi_n F_{n+1},\\
\ddot q_N=F_{N-1}-F_N(1+\varphi_{N-1}).
\end{split}
\end{equation}
For simplicity, we assume that the quantities $\omega_i$ and $m_i$
are the geometric sequences. So, $\omega_i=\omega_0 r^{i-1}$,
$r>1$, and $m_i=m_0 h^{i-1}$, $h<1$. From this it follows that
$\varphi_i=sh=\varphi=\mbox{const}$.

\section{Studies of the model with three hierarchical
levels}\label{sec:2}

To begin with, let us consider  the model with three hierarchical
levels, i.e. putting $N=3$. In this case system (\ref{syst1}) has
the following form \begin{equation}\label{syst3}
\begin{split}
\ddot q_1=-F_1+\varphi F_2,\\
\ddot q_2=F_1-F_2(1+\varphi)+\varphi F_3,\\
\ddot q_3=F_2-F_3(1+\varphi).
\end{split}
\end{equation}
Till now we did not achieve success in finding any exact solutions
to system (\ref{syst3}). Therefore, in order to get the
information on the solutions structure, we are going to use the
numerical and qualitative analysis methods.

 Due to the Hamiltonian nature of  system
(\ref{syst3}), the trajectories of the dynamical system lay on
constant energy hyper-surfaces. The allocation of these
hyper-surfaces in the phase space can be effectively studied by
means of the Poincar\'{e} section technique
\cite{holodn,workbook}.

\subsection{The sequence of frequencies $\omega_i=r^{i-1}$,
$r=1.1$}\label{subsec_f11}

\begin{figure*}[tbh]
\includegraphics[width=8 cm, height=6 cm ]{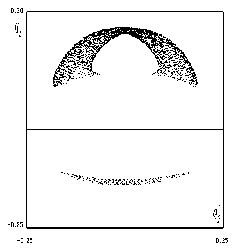}
\hfill
\includegraphics[width=8 cm, height=6 cm ]{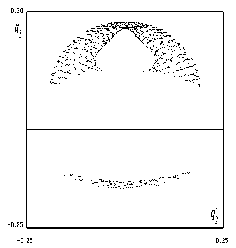}
 \centerline{a \hspace{8cm} b}\\
\includegraphics[width=8 cm, height=6 cm ]{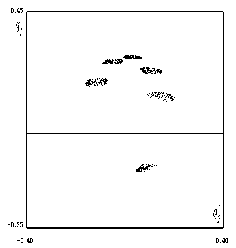}
\hfill
\includegraphics[width=8 cm, height=6 cm ]{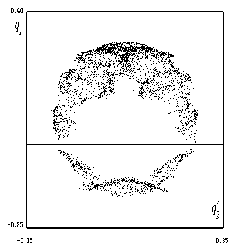}
 \centerline{c \hspace{8cm} d}
\caption{Poincar\'{e} sections at different $\beta=1$ (a),
$\beta=1.05$ (b), $\beta=1.10$ (c),$\beta=1.15$ (d). Initial
conditions $q_1^\prime = 0.2$ and the other coordinates are equal
to zero.}\label{Poinc:1}
\end{figure*}

Let us fix the parameters $\omega_0 = 1$, $r=1.1$, $\varphi=0.6$
and choose the initial conditions for numerical integration in the
form of $q_1(0)=0$, $q_1^\prime(0)=0.2$, $q_j(0)=q_j^\prime
(0)=0$, $j=2,3$. At first, consider the linear case when
$\beta=1$. Let the plane $q_1^\prime=0$ be the Poincar\'{e}
section plane. Integrating of system (\ref{model_ODE2}) by means
of the Dormand-Prince method \cite{DoPri}, we are interested in
the points in which trajectories intersect the cross-plane in one
direction only. Mapping this section onto the plane
$(q_2^\prime;q_1)$, we obtain the diagrams presented in the figure
\ref{Poinc:1}a. The figure \ref{Poinc:1}b corresponds to the
Poincar\'{e} section of the trajectory beginning at the same
starting point as in previous case but the parameter $\beta=1.05$.

\begin{figure*}[t]
\begin{center}
\includegraphics[width=7 cm, height=5 cm ]{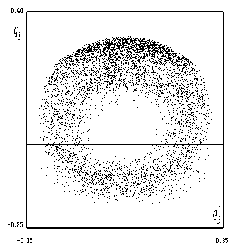}
\hspace{1cm}
\includegraphics[width=7 cm, height=5 cm ]{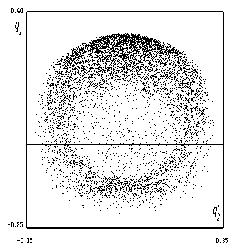}
 \centerline{a \hspace{8cm} b}\\
\includegraphics[width=7 cm, height=5 cm ]{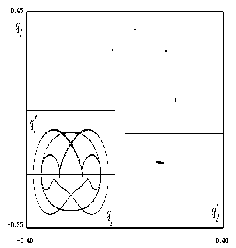}
\hspace{1cm}
\includegraphics[width=7 cm, height=5 cm ]{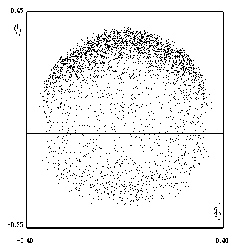}
 \centerline{c \hspace{8cm} d}
\end{center}
\caption{Poincar\'{e} sections at  $\beta=1.20$ (a), $\beta=1.25$
(b), $\beta=1.5$ (c), (d). Initial conditions for (a), (b), (c)
$q_1^\prime= 0.2$ and other coordinates are equal to zero. Initial
condition for (d): $q_1=0.068$, $q_1^\prime = 0.2$,
$q_{2,3}=q_{2,3}^\prime =0$. In the corner of the figure (c) the
quasi-periodic trajectory's phase portrait corresponding to the
Poincar\'{e} section (c) is drown. }\label{Poinc:2}
\end{figure*}

In the same manner, putting $\beta =1.1$, $\beta =1.15$, $\beta
=1.25$, and $\beta =1.5$ (the nonlinear cases), we obtain the
typical Poincar\'{e} diagrams depicted in the
Figs.\ref{Poinc:1}-\ref{Poinc:2}.

From the analysis of these figures it follows that the
incorporation of nonlinearity in the model causes the appearance
of the striped torus surfaces (Figs.\ref{Poinc:1}b), tori with
dense winding (Figs.\ref{Poinc:1}c), trajectories similar to
periodic ones (Figs.\ref{Poinc:2}c), which are not observed in the
linear case.

According to  Fig.\ref{Poinc:2}d, the increasing of nonlinearity
associated with the parameter $\beta$ leads to the formation of
Poincar\'{e} sections uniformly filled with points.

Comparing the initial conditions for construction of the
Poincar\'{e} diagrams of  Fig.\ref{Poinc:2}, c and d, we see that
small deviation of the first coordinate changes the observed
regimes essentially. This is one of the main feature of the
nonlinear models \cite{Ott}.

It worth noting that the sizes of regimes differ from each other
weakly. But there are values of the parameter $\beta$ when the
region  of the Poincar\'{e} section (Fig.\ref{Poinc:2}a,b,d) is
filled with points nonuniformly. This tells us about existence of
prevailing amplitudes in the quasi-periodic regime.

Additional information on the properties of regimes we study can
be obtain with the help of the correlation analysis. Using the
solution of system (\ref{model_ODE2}) in the form of the discrete
sequences $q_i(j)$, $i=1,2,3$, $j=1\cdot \tau, 2\cdot \tau\ldots,
M\cdot \tau$ ($\tau=0.01$ is the step of the temporal variable
discretization), we define the cross-correlation function
\cite{autocorr}
$$
R_{xy}(k)=\frac{M}{M-k}\frac{\sum_{j=1}^{M-k}q_x(j)q_y(j-k)}{\sum_{j=1}^{M}q_x(j)q_y(j)},\quad
{x,y}=1,\ldots,3.
$$
If $x=y$, then we get the definition of the autocorrelation
function $ R_{xx}(k)$. Together with the correlation functions we
use their Fourier transformations
$\mbox{FFT}(R_{xy})=S_{xy}(\omega)$ which, in the case of
autocorrelation function, coincide with the power spectrum of a
signal \cite{for_autocor}.

\begin{figure}
\begin{center}
\includegraphics[width=7 cm, height=5 cm ]{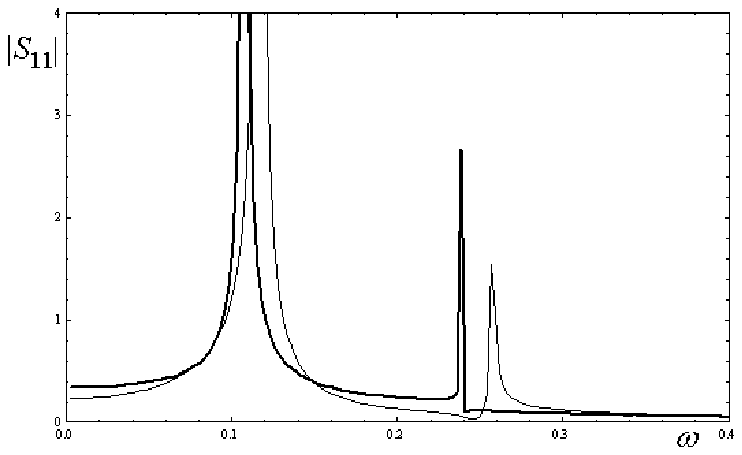}
\hspace{1cm}
\includegraphics[width=7 cm, height=5 cm ]{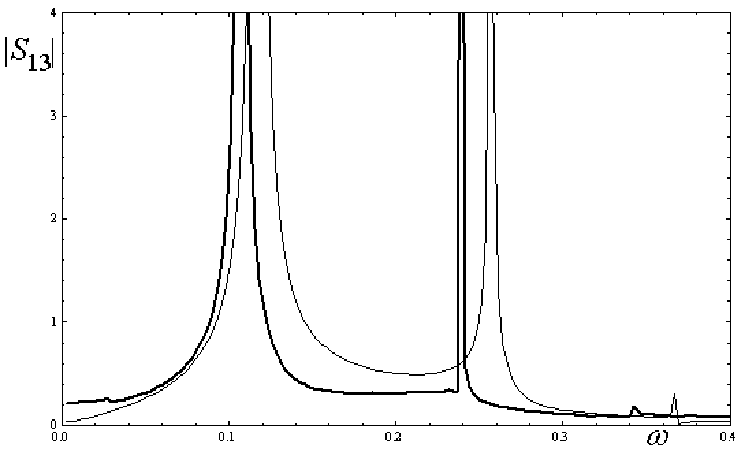}
 \centerline{a \hspace{8cm} b}\\
 \includegraphics[width=7 cm, height=5 cm ]{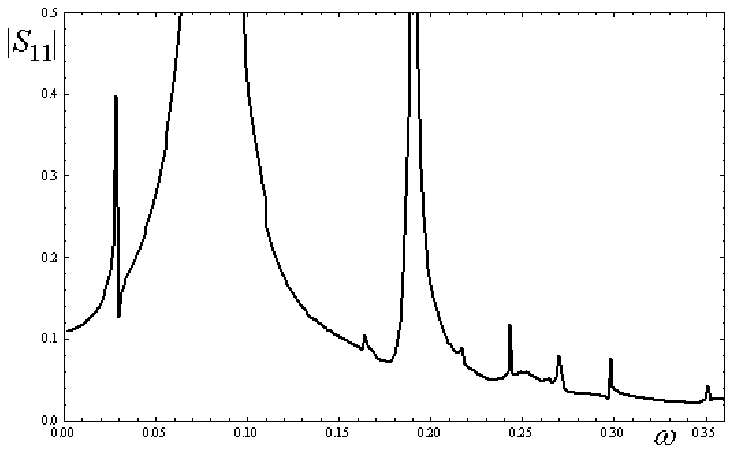}
\hspace{1cm}
\includegraphics[width=7 cm, height=5 cm ]{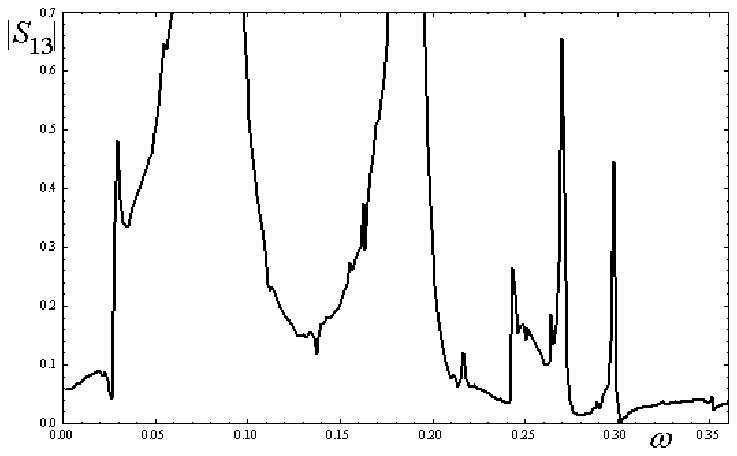}
 \centerline{c \hspace{8cm} d}
 \end{center}
\caption{The graphs of the functions $|S_{11}(\omega)|$ (left) and
$|S_{13}(\omega)|$ (right). The spectra correspond to the regime
of Fig.~\ref{Poinc:1}c (top panels) and Fig.~\ref{Poinc:2}d
(bottom panels). The curves depicted by  thin lines correspond to
linear system (\ref{model_ODE2}) with $\beta=1$. }\label{Autocorr}
\end{figure}

Let us begin from the correlation analysis of regimes derived at
small $\beta$. Taking the trajectories whose Poincar\'{e} sections
are depicted in  Figs.\ref{Poinc:1}a,c, we obtain the graph of
$|S_{11}(\omega)|$ (Fig.\ref{Autocorr}a) and $|S_{13}(\omega)|$
(Fig.~\ref{Autocorr}b).  According to  Fig.~\ref{Autocorr}a, the
spectrum has the essential maximum near $\omega=0.1$ and the other
one at about $2\omega$. This tells us that  the sequence $q_1$
possesses the prevailing temporal scale when correlation is the
most strong. Note that this dominant frequency displaces to the
left  at growing nonlinearity (to compare thick and thin curves).
Since $\beta$ is small, oscillators' behavior does not differ from
each other (Fig.~\ref{Autocorr}b).

For  $\beta=1.5$ when chaotic regime exists (Fig.\ref{Poinc:2}d),
the spectrum of the function $R_{11}$ has two multiple prevailing
frequencies (Fig.\ref{Autocorr}c) which are  accompanied by
several peaks. At the same time, we observe more essential
 defragmentation of  spectrum for the function
$R_{13}$ (Fig.\ref{Autocorr}d).  Especially, it should be paid
attention to the appearance of additional high-frequency peaks
which can be regarded as  supplementary temporal scales in media.

\subsection{The sequence of frequencies $\omega_i=r^{i-1}$,
$r=2.0$}

Now we are interested in the studies of  oscillations in system
(\ref{model_ODE2}) when the partial frequencies are following
$\omega_i=r^{i-1}$, $r=2.0$. We keep the same initial conditions
for numerical integration as in the previous subsection
\ref{subsec_f11}.

\begin{figure}
\includegraphics[width=8 cm, height=6 cm ]{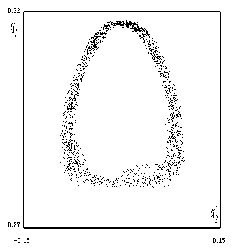}
\hfill 
\includegraphics[width=8 cm, height=6 cm ]{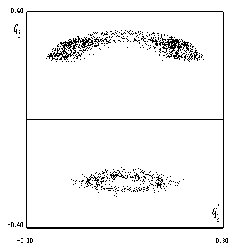}
 \centerline{a \hspace{8cm} b}\\
\includegraphics[width=8 cm, height=6 cm ]{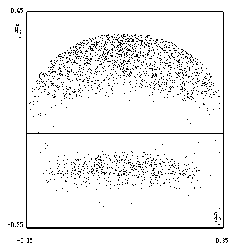}
\hfill 
\includegraphics[width=8 cm, height=6 cm ]{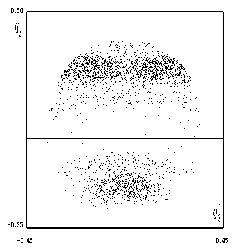}
 \centerline{c \hspace{8 cm} d}
 \caption{Poincar\'{e} sections at  $\beta=1.2$ (a),
$\beta=1.3$ (b), $\beta=1.4$ (c), $\beta=1.5$ (d). The values of
parameters $r=2$, $\varphi=0.6$. Initial conditions $q_1^\prime=
0.2$ and the other coordinates are equal to zero. }\label{Poinc:3}
\end{figure}

From the Poincar\'{e} sections (Fig.\ref{Poinc:3}a) it follows
that the trajectories can form the pipe in the phase space at
small values of parameter $\beta$. When $\beta$ increases, we
observe the transformation of phase space in such a way that the
regimes become more chaotic (Fig. \ref{Poinc:3}b,c). Finally, the
growth of $\beta$ leads to essential chaotization of the
Poincar\'{e} section (Fig.\ref{Poinc:3}d).

The growth of $\beta$ is thus  accompanied by increasing of
irregularity. In this situation we can not reduce system
(\ref{model_ODE2})  to low dimensional model like one dimensional
maps generated by Poincar\'{e} sections.

\begin{figure}
\includegraphics[width=5 cm, height=4 cm ]{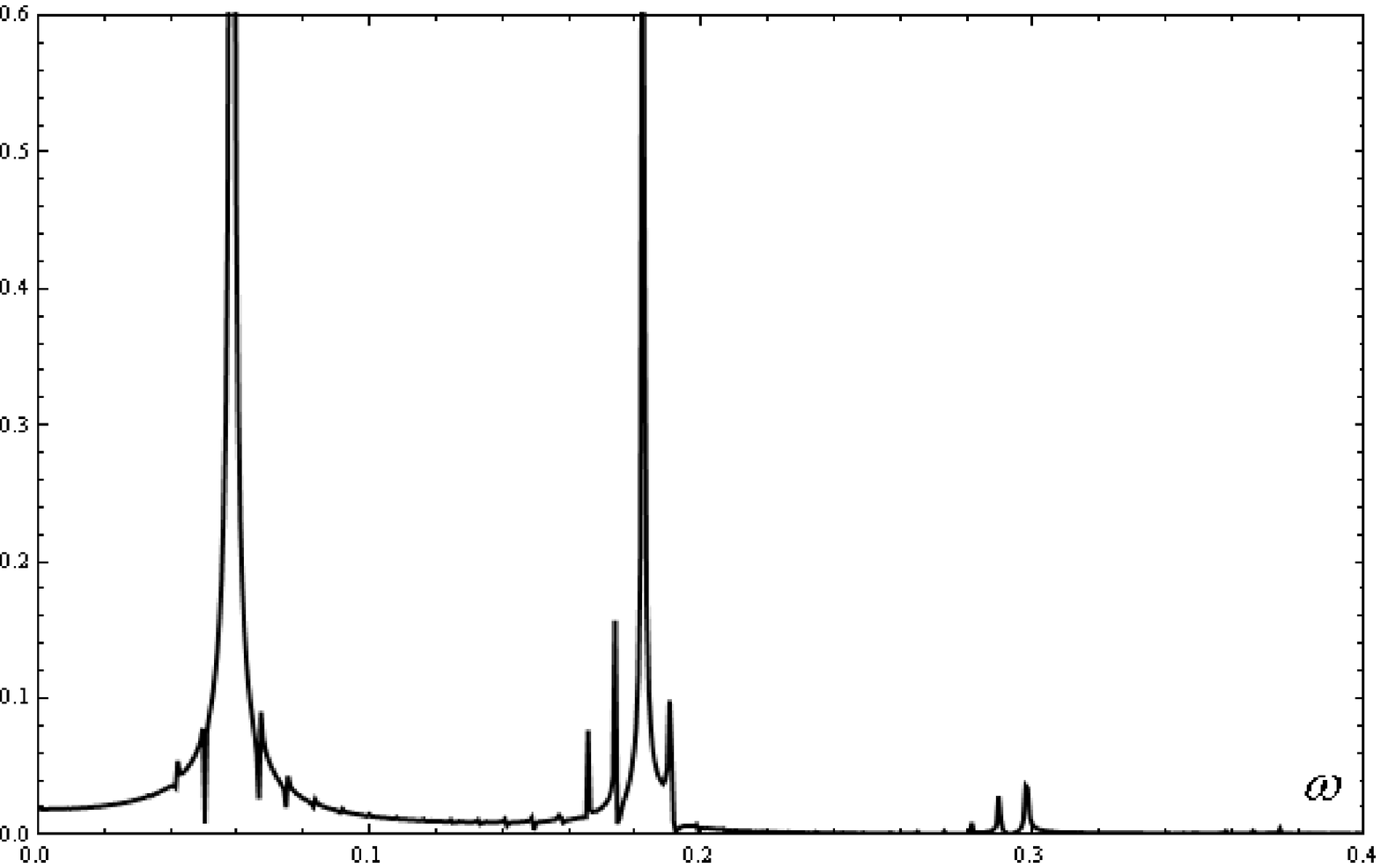} \includegraphics[width=5 cm, height=4 cm ]{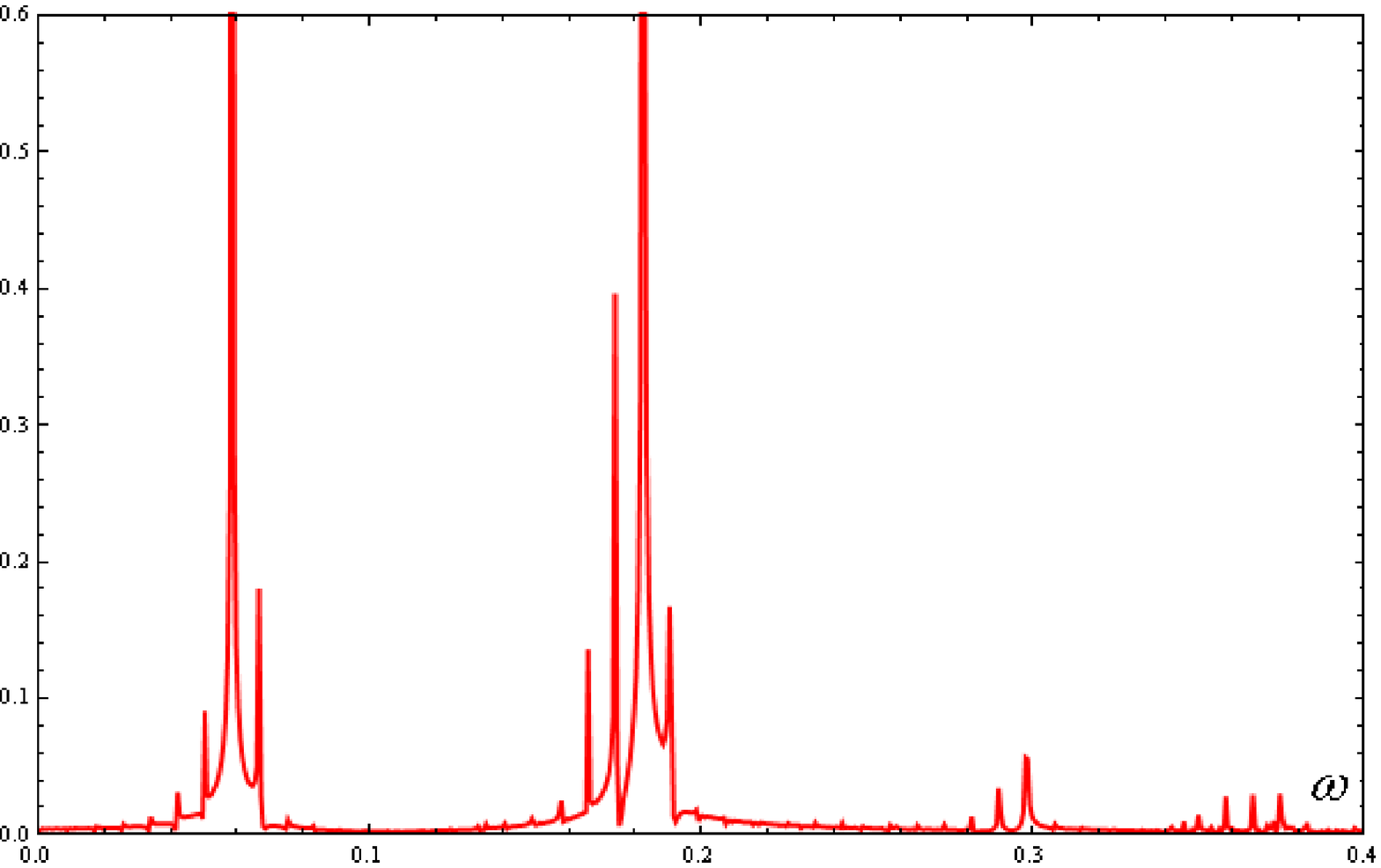}
\includegraphics[width=5 cm, height=4 cm ]{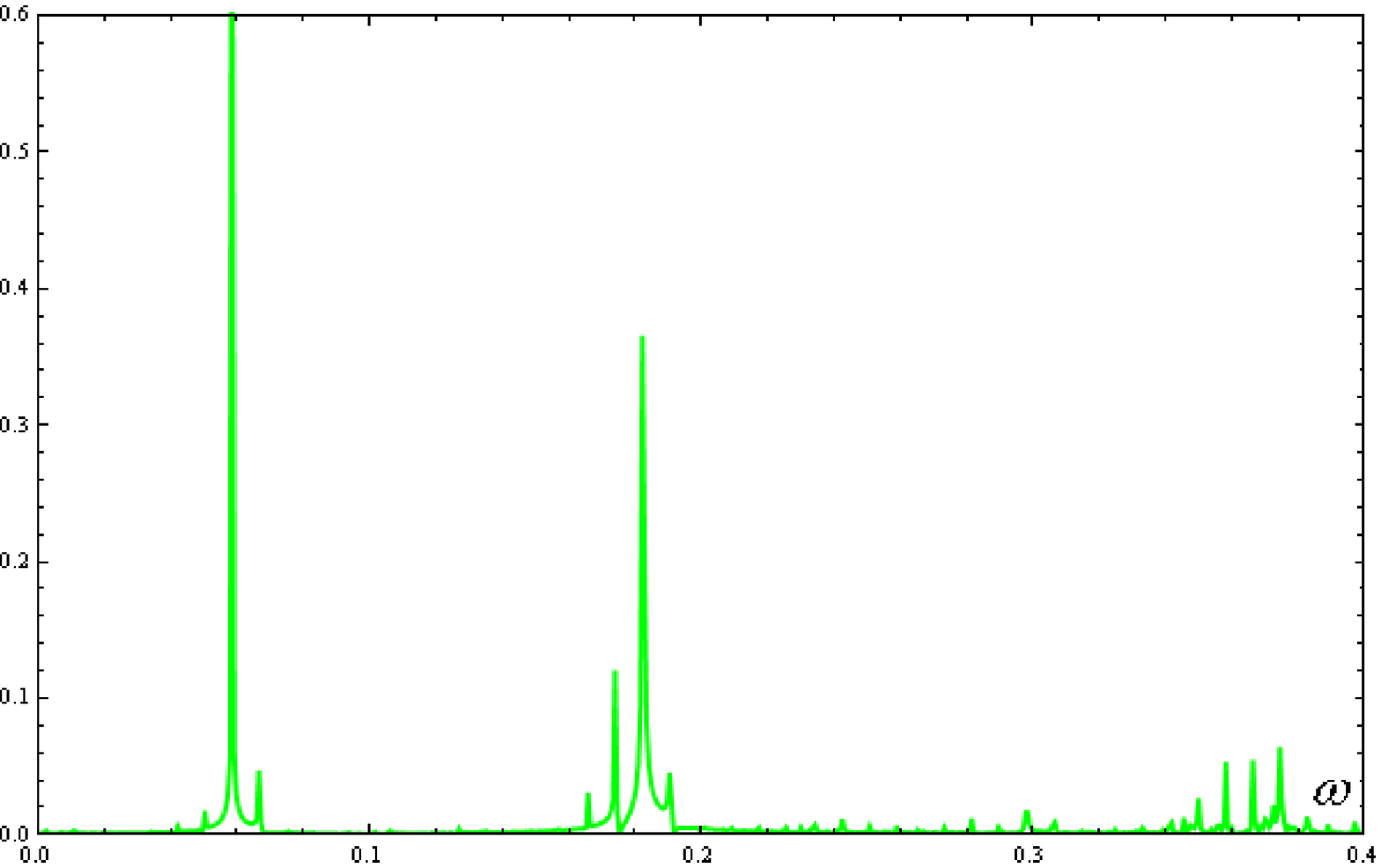}
\hfill
 \centerline{a \hspace{5cm} b \hspace{5cm} c}
\includegraphics[width=5 cm, height=4 cm ]{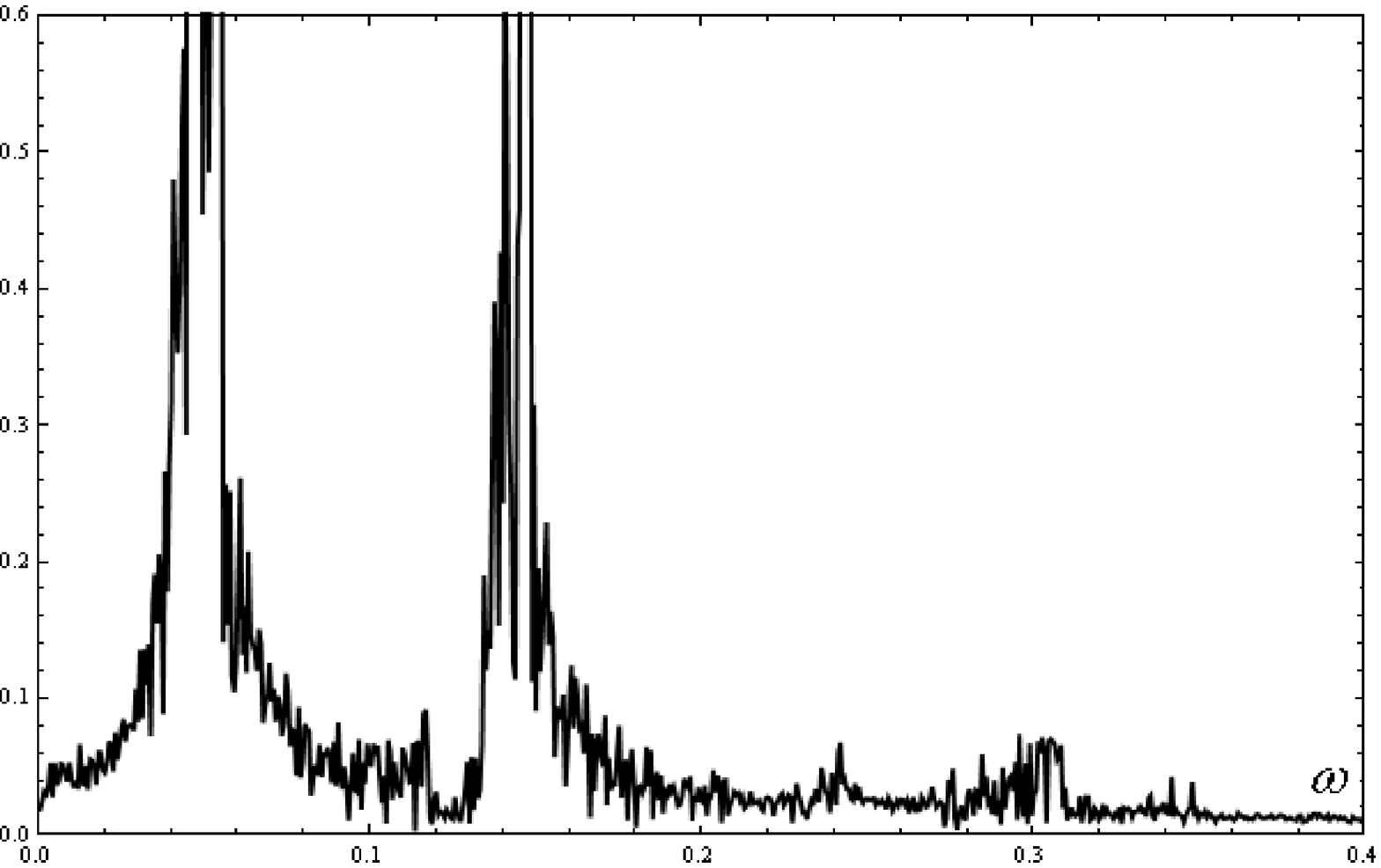} \includegraphics[width=5 cm, height=4 cm ]{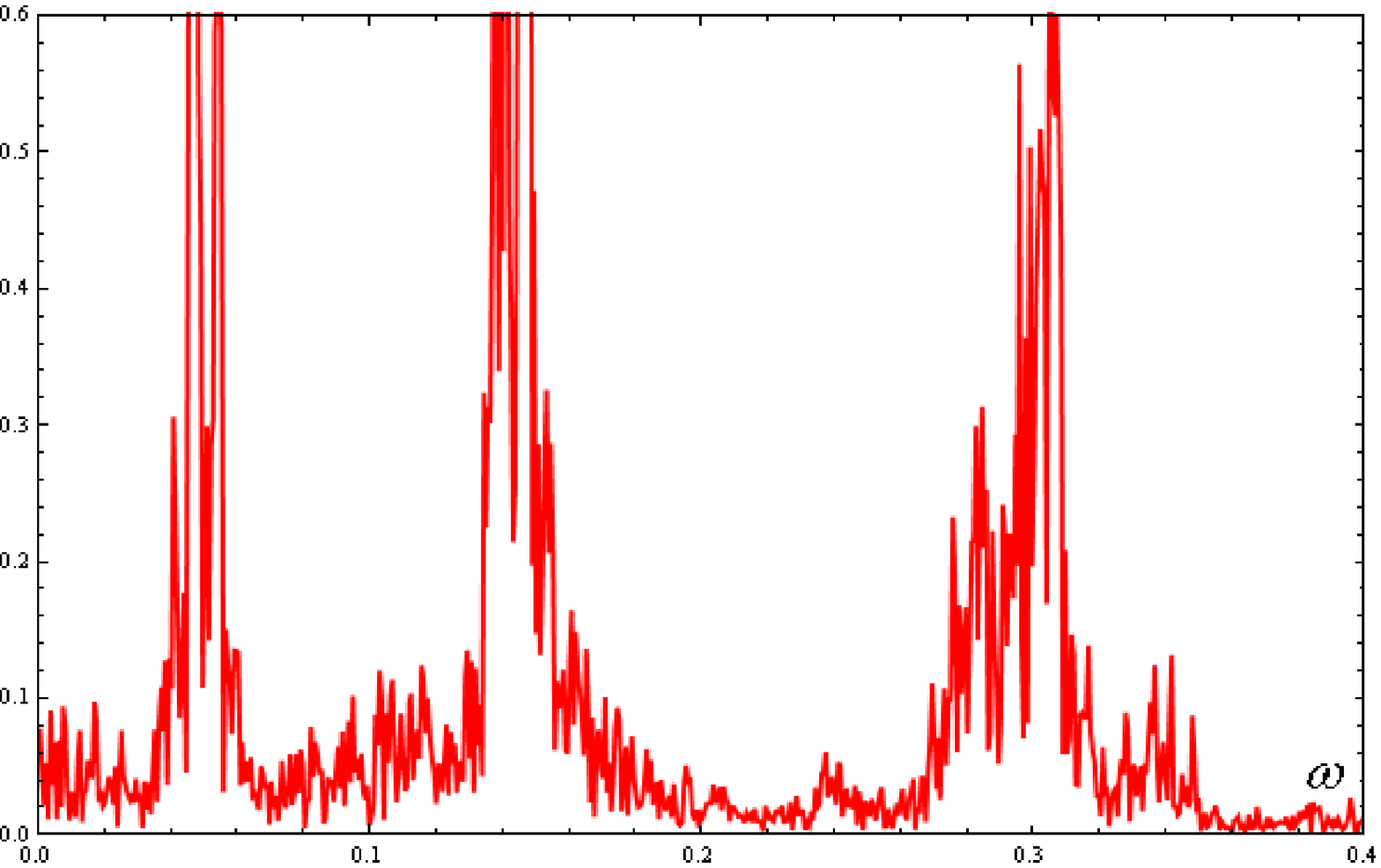}
\includegraphics[width=5 cm, height=4 cm ]{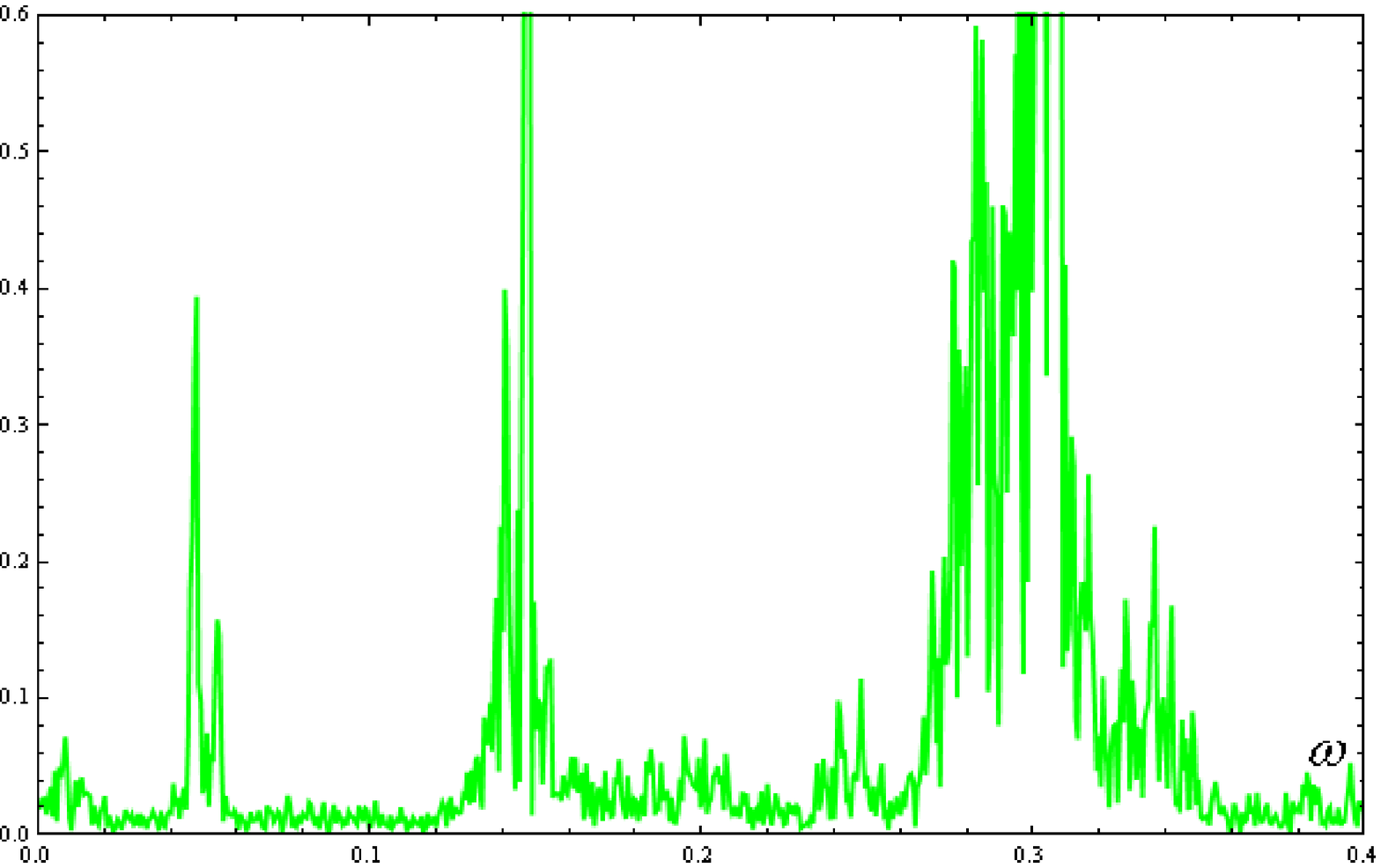}
 \centerline{d \hspace{5cm} e \hspace{5cm} f}
\caption{(color online) Fourier spectra of the $q_i$,
$i=1,2,3$-component of the trajectories drawn in the
Fig.~\ref{Poinc:3} at $\beta=1.2$ and $\beta=1.5$, respectively.
}\label{Fourier}
\end{figure}

To understand phenomena in the system at increasing $\beta$, we
derive the Fourier transformations for components $q_{i}$, namely,
$FFT(q_i)$ at $\beta=1.2$ (Fig.\ref{Fourier}a-c) and $FFT(q_i)$ at
$\beta=1.5$  (Fig.\ref{Fourier}d-f). From spectra presented in
Fig.\ref{Fourier}a-c it follows that all components $q_i$ are
characterized by two essential spectral  maxima. This  causes the
existence of torus (Fig.\ref{Poinc:3}a) in  system's phase space.

For $\beta=1.5$, the component $q_1$ has two wide spectral maxima
(Fig.\ref{Fourier}d) and one notable maximum in the same place as
in the spectrum from Fig.\ref{Fourier}a.

The spectrum of $q_2$ (Fig.\ref{Fourier}e) involves  three
prevailing maxima having strongly irregular character. In the
spectrum of $q_3$ we can distinguish three maxima as well. In
contrast to the spectrum from Fig.\ref{Fourier}c, we see that the
height of maxima grows as $\omega$ increases. This can be
associated with the excitation of the high-frequency spectral
modes.

Therefore, system (\ref{model_ODE2}) can be regarded as a model
describing  the directed transport of energy in hierarchical
media.



\section{Studies of the model with many hierarchical levels}\label{sec:3}

\begin{figure}
\begin{center}
\includegraphics[width=6.5 cm, height=5 cm ]{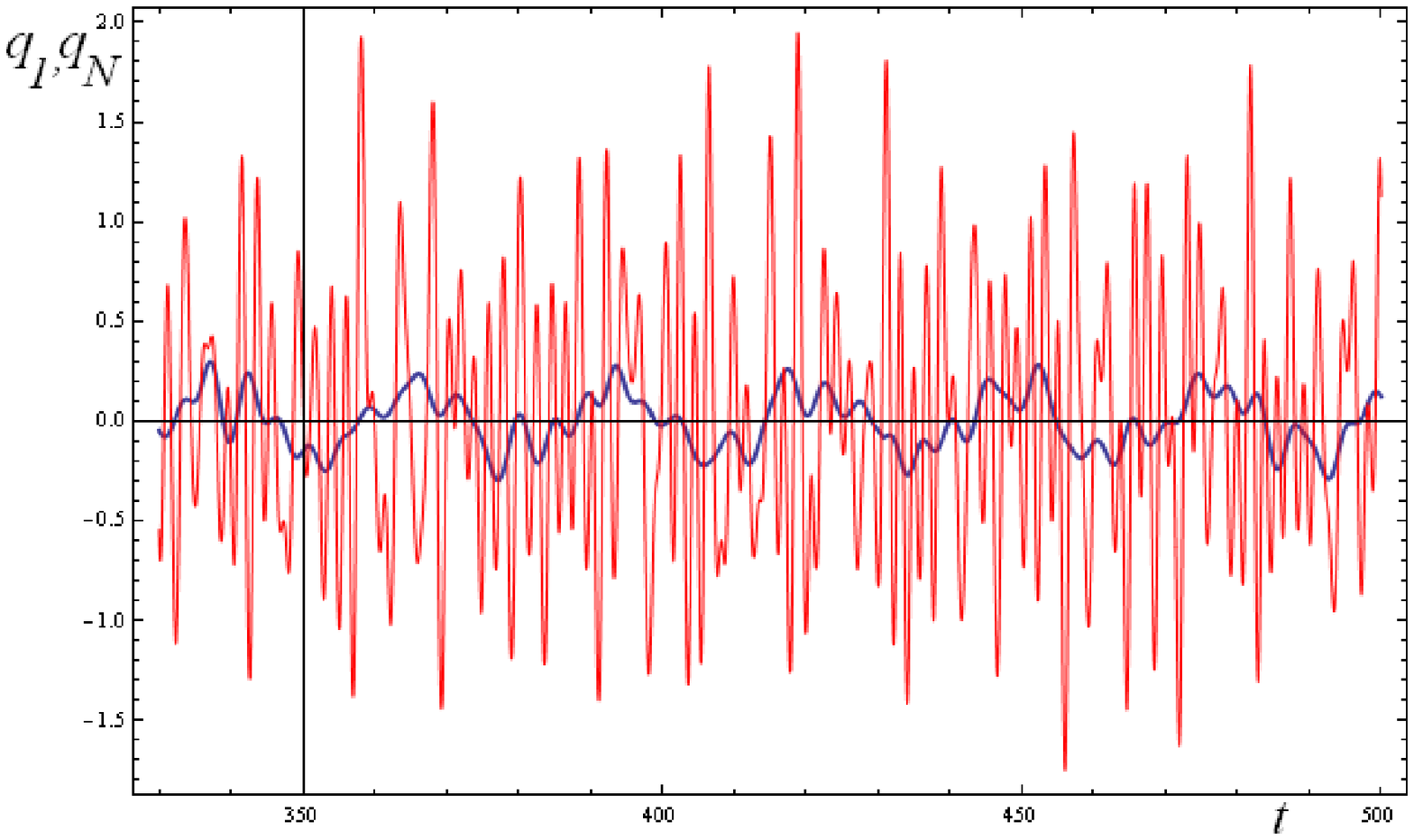}
\hspace{1 cm}
\includegraphics[width=6.5 cm, height=5 cm ]{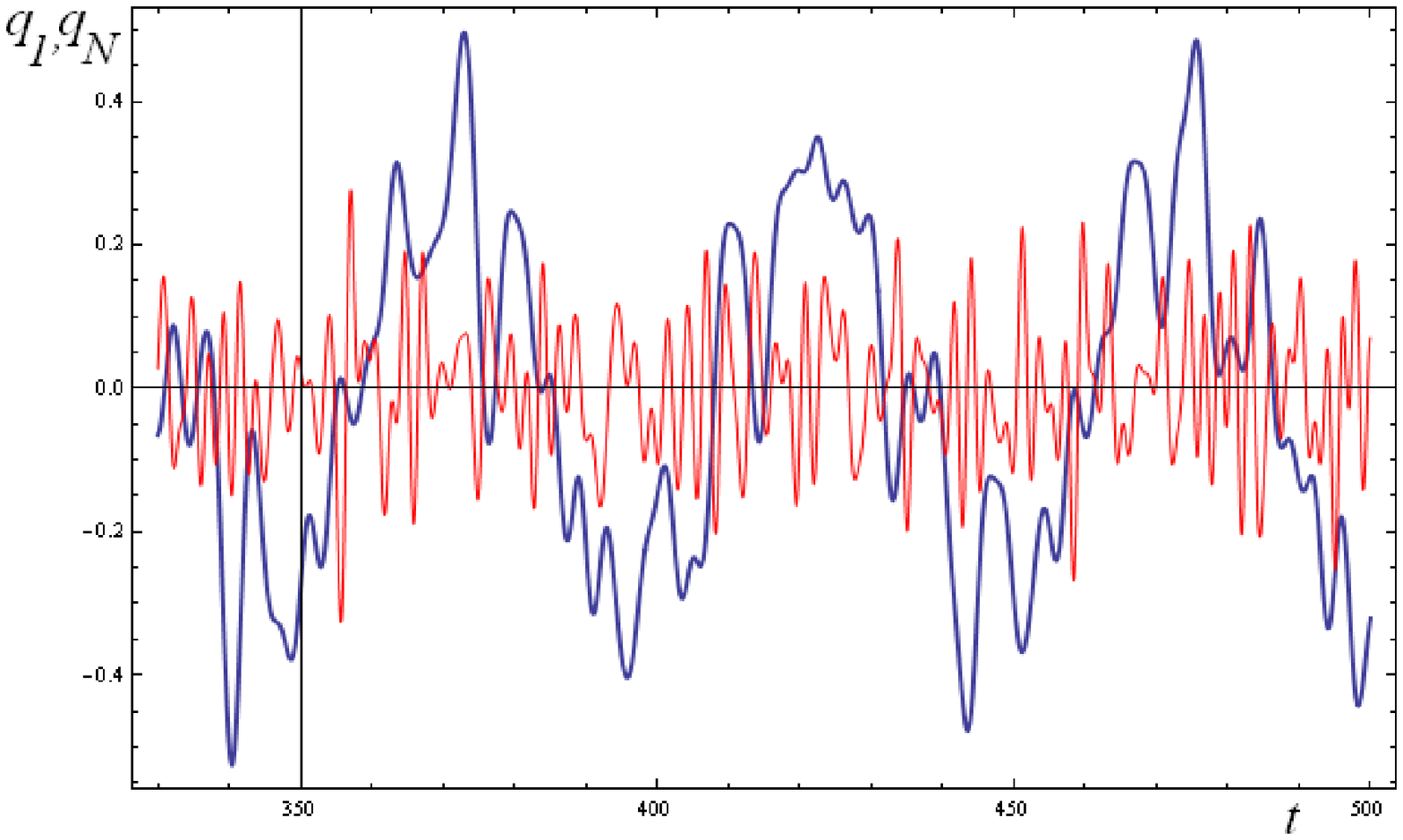}\\
 a \hspace{7 cm} b\\
\end{center}
\caption{(color online) Snapshots  of the components $q_1(t)$
(thick curve) and $q_N(t)$ (thin red curve), $t\in (330;500)$ at
$\varphi=0.7$ (a) and $\varphi=0.9$ (b).
 }\label{rozgortka}
\end{figure}

\begin{figure}
\begin{center}
\includegraphics[width=6.5 cm, height=5 cm ]{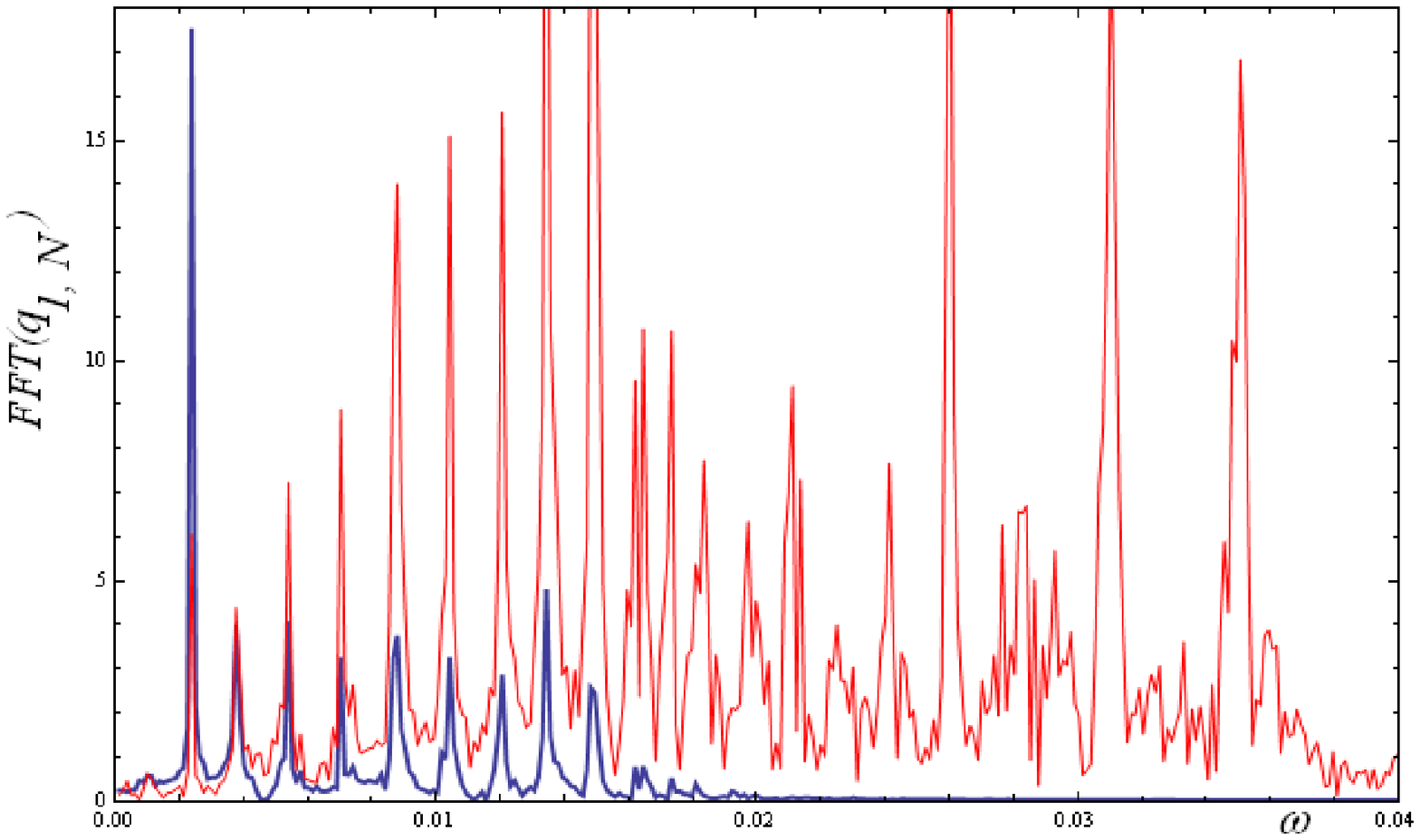}
\hspace{1 cm}
\includegraphics[width=6.5 cm, height=5 cm ]{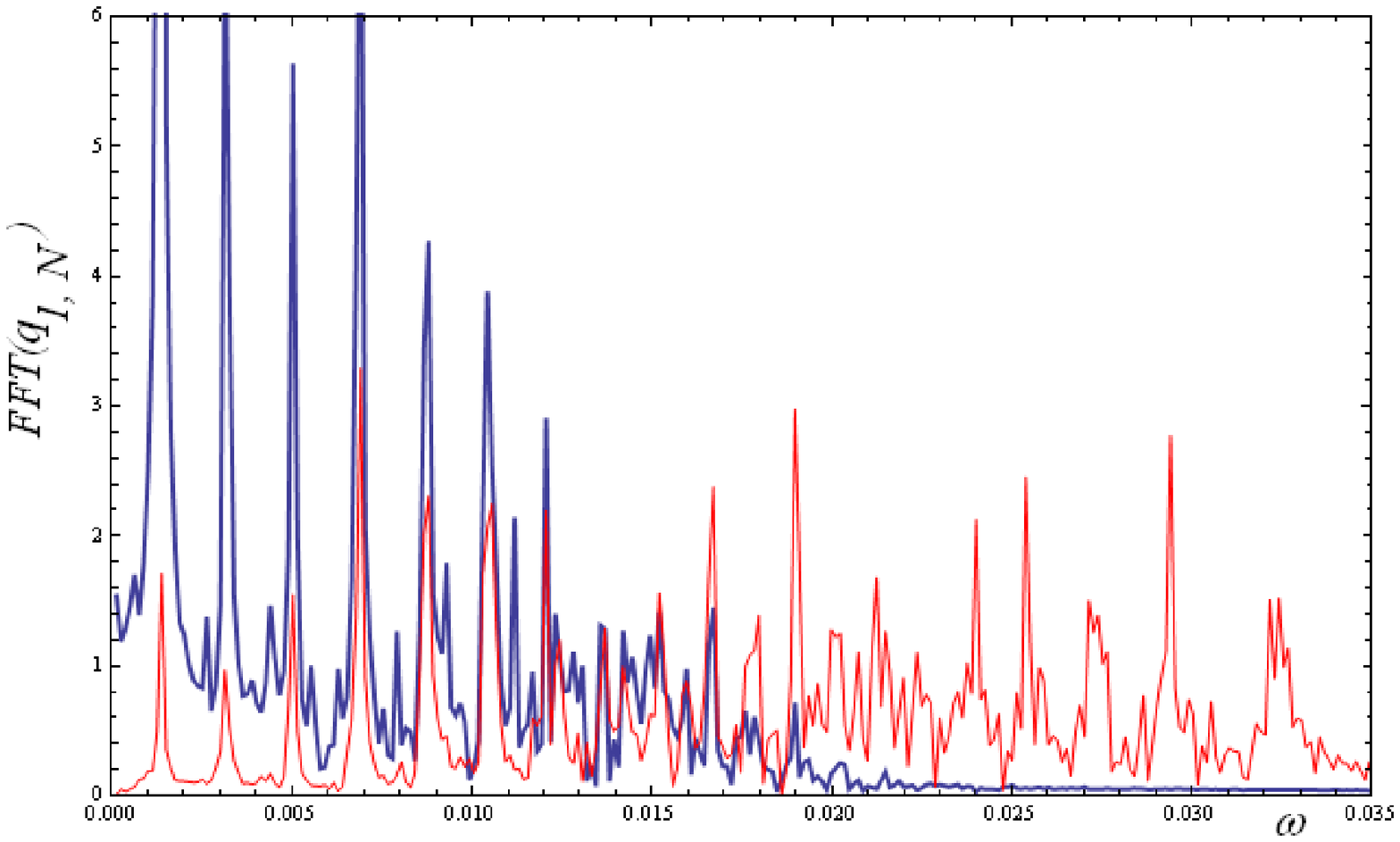}\\
a \hspace{7 cm} b\\
\end{center}
 \caption{(color online) Fourier spectra for the components $q_1(t)$ (thick
curve) and $q_N(t)$ (thin red curve) at $\varphi=0.7$ (a) and
$\varphi=0.9$ (b).
 }\label{fourier20}
\end{figure}

Consider now the model which consists of many  hierarchical
levels, for instance, $N=20$. The fixed parameters are as follows
 $\omega_0=1$, $r=1.04$, $\beta=1.15$ and initial
condition $q_1(0)=0$, $q_1^\prime(0)=0.3$, $q_j(0)=q_j^\prime
(0)=0$, $j=2,\ldots, N$. Compare the $q_1$ and $q_N$ components of
solutions to system (\ref{model_ODE2}) derived at $\varphi=0.7$
and $\varphi=0.9$. Corresponding results are presented in the left
and right panels of  figures \ref{rozgortka} and \ref{fourier20},
respectively.

For $\varphi=0.7$ the amplitude of $q_1$ is much less than the
amplitude of $q_N$. Snapshot from Fig.\ref{rozgortka}a predicts
that frequency anatomy of the $q_1$ and $q_N$ signals is
different. Moreover, the $q_1$ dynamics looks like a harmonic
signal modulated by another signal with higher frequency. This is
confirmed by the analysis of Fourier spectra for $q_1$ and $q_N$
(Fig.\ref{fourier20}a). The spectrum for $q_1$ depicted by thick
line in Fig.\ref{fourier20}a contains one maximum corresponding to
existence of mode like a cosine function and a number of much
smaller maxima causing  a ripple on the $q_1$ profile.

As Fig.\ref{fourier20}a testifies, the spectrum for $q_N$ depicted
by the red thin curve contains a number of almost equivalent
maxima located both in the low and  high frequency zones.

Similar analysis can be carried out for the solutions to system
(\ref{model_ODE2}) at $\varphi=0.9$. Snapshots for the components
$q_1$ and $q_N$ plotted in Fig.\ref{rozgortka}b show that the
amplitude of $q_1$ is larger than $q_N$. According to
Fig.\ref{fourier20}b, the main part of $q_1$ Fourier spectrum is
localized in low frequency zone. But the spectrum of $q_N$ is
almost uniformly distributed in the domain under consideration.

From the results presented in this section we can  conclude that,
at first, a critical value of $\varphi$ can exist corresponding to
the formation of comparable oscillations on the first and the last
 hierarchical levels of media. Secondly, the spectrum of the
lowest level we are most interested in  is distributed in a wide
frequency domain and depending on the $\varphi$ dominant
frequencies can be distinguished.

\section{Conclusions}\label{sec:4}

In summary, the strongly nonlinear system of coupled oscillators
describing media with hierarchical structure is introduced. The
problem we have stated is concerned with the peculiarities of
energy transfer along the hierarchical structure of media. To
treat this problem, we have studied the  phase portraits,
correlation functions and Fourier spectra which characterize
oscillators placed on different hierarchical levels of media. We
thus have seen that  hierarchical systems manifest quasiperiodic
and chaotic regimes development  of  which depends on the
auxiliary parameter $\varphi$.  It is shown that among values of
the parameter   $\varphi$ corresponding to processes of
transferring energy from the top level to the lowest one there is
a threshold value. Therefore, within the framework of presented
model, one can confirm that the hierarchical structure accompanied
by  nonlinearity   plays an important role in the transformation
of energy flows in media. Rearrangement of the hierarchical levels
caused by intensive loading is the natural mechanism of
accumulation  and radiation of energy in structured media under
strongly non-equilibrium conditions.

\end{document}